\documentclass[reprint, superscriptaddress, nofootinbib, amsmath, amssymb, amsfonts, aps, pra]{revtex4-1}

\usepackage{graphicx}
\graphicspath{ {/paper/}}
\usepackage{dcolumn}
\usepackage{amsthm}
\usepackage{bm}
\usepackage[english]{babel}
\usepackage{slashed}
\usepackage{enumitem}
\usepackage{blindtext} 
\usepackage[linktocpage]{hyperref}
\usepackage[svgnames]{xcolor}
\hypersetup{
     colorlinks = true,
     linkcolor = Maroon,
     anchorcolor = black,
     citecolor = DarkBlue,
     filecolor = blue,
     urlcolor = DarkBlue
     }

\usepackage{bbold}
\usepackage{braket}
\usepackage{tikz}
\usetikzlibrary{quantikz}
\usetikzlibrary{positioning}
\usepackage{kantlipsum}

\newtheorem{theorem}{Theorem}
\newtheorem{corollary}{Corollary}

\newtheorem{proposition}{Proposition}

\newcommand{\revise}[1]{\textcolor{black}{#1}}

\usetikzlibrary{shadows,arrows,calc,backgrounds,fit,shapes,snakes,shapes.multipart,decorations.pathreplacing,shapes.misc,patterns,positioning}
\usepackage{pgfplots}
\usetikzlibrary{backgrounds,fit,decorations.pathreplacing}

\begin{document}


\title{On the emergence of quantum memory in non-Markovian dynamics}

\author{Alexander Yosifov}
\email{alexanderyyosifov@gmail.com}
\affiliation{School of Physical and Chemical Sciences, Queen Mary University of London, London, E1 4NS, United Kingdom}

\author{Aditya Iyer}
\email{aditya.iyer@physics.ox.ac.uk}
\affiliation{Clarendon Laboratory, University of Oxford, Parks Road, Oxford, OX1 3PU, United Kingdom}

\author{Vlatko Vedral}
\email{vlatko.vedral@physics.ox.ac.uk}
\affiliation{Clarendon Laboratory, University of Oxford, Parks Road, Oxford, OX1 3PU, United Kingdom}

\author{Jinzhao Sun}
\email{jinzhao.sun.phys@gmail.com}
\affiliation{School of Physical and Chemical Sciences, Queen Mary University of London, London, E1 4NS, United Kingdom}

\date{\today}

\begin{abstract}

The emergence of memory is a hallmark feature of non-Markovian dynamics. However, the type of memory---classical or quantum---required to realize certain dynamics remains unknown. We study the quantum homogenizer as a minimal model of non-Markovian evolution and identify the physical conditions under which genuinely quantum memory becomes necessary. Using entanglement measures and relying only on the local dynamics as a witness, we prove both analytically and numerically the type of memory depends not merely on the dynamics itself, but also on the reservoir's initial entanglement structure, and in particular the propagation of non-classical correlations within it. For different bi- or multi-partite reservoir initializations, we establish a correspondence between interaction strength and entanglement generation. We provide physical criteria and an activation lower bound for the onset of quantum memory. The results may inform us how environmental correlations govern the transition from classical to quantum memory in open quantum systems.

\end{abstract}

\maketitle

\let\oldaddcontentsline\addcontentsline
\renewcommand{\addcontentsline}[3]{}

\revise{
\textit{Introduction---}Non-Markovianity is ubiquitous in realistic quantum systems \cite{3,4} and has recently emerged as an important resource for enhancing quantum control in solid-state and NMR devices \cite{qcontrol1, qcontrol2, qcontrol3}, improving quantum communication protocols \cite{qcomm}, and mitigating off-resonant gate errors \cite{NMhelp}. The characteristic microscopic interactions with the environment, however, preserve correlations over time, allowing past information to perpetuate and influence future dynamics, hence giving rise to \textit{memory}. While the involvement of a large number of environmental degrees of freedom makes such dynamics inherently difficult to simulate, methods like the quantum hierarchical equations of motion \cite{hardd}, the time-evolving matrix product operator \cite{hardd2}, and the process tensor \cite{hardd3} have made significant progress in tackling that.}

\revise{
Still, a key remaining challenge lies in the ambiguous nature of the memory itself. Although all non-Markovian processes exhibit some form of memory effects, those need not necessarily be genuinely quantum; and in fact, in many cases classical memory suffices \cite{memory1, memory2, memory3, memory4}. Distinguishing between classical and quantum memory is crucial, as the type can have important consequences for the design, implementation, and performance of noisy quantum devices \cite{add1, add2, add3, add4}, with applications to continuous quantum error correction (QEC) \cite{oo}, coherence retention in superconducting processors \cite{coherence}, quantum teleportation \cite{qteleport}, quantum key distribution \cite{qkd}, and quantum feedback control \cite{control}. }

\revise{
Nevertheless, the onset of memory in non-Markovian dynamics has so far been discussed primarily from a computational perspective, without clear understanding of the underlying mechanisms that dictate it. Meanwhile, existing definitions of non-Markovianity, like those based on dynamical map divisibility or the trace distance \cite{criteria1,criteria2}, are useful only in detecting the presence of non-Markovian dynamics, but remain agnostic as to the nature of the memory. Therefore, the question of what physical conditions (system-reservoir interaction strength and reservoir state initialization) require genuine quantumness of the memory remains open. One way to address it is by identifying the threshold at which quantum memory becomes indispensable. }

\revise{
In this work, we consider the quantum homogenizer \cite{maria,ziman,vlatko,alexander2,alexander}---a minimal, yet representative, model of non-Markovian dynamics---and study the optimal entanglement transfer conditions between the system and the reservoir. As a memory witness, we use the criterion established in \cite{memory} which is based solely on the local dynamics, with concurrence as the entanglement measure. We find that determining the type of memory is subtle, as it depends not only on the dynamics itself but also on the reservoir’s initial entanglement structure, and more specifically the propagation of non-classical correlations within it. For different classes of reservoir initializations, we establish, both numerically and analytically, a correspondence between the microscopic coupling strength and the rate at which entanglement is generated, and prove this growth is upper bounded by the Bell-state marginal. We further study a more general, multipartite scenario that involves an effective intermediate decoherence channel suppressing the quantum correlations in the reservoir \cite{asymm1,asymm2,asymm3}, and examine the role of higher-order effects. Our work establishes the physical criteria for the onset of quantum memory in non-Markovian dynamics, identifying the minimal interaction strength and reservoir correlations required; which can inform when standard Markovian error-mitigation techniques become invalid. Moreover, in solid-state and exchange-based models, where partial-\texttt{SWAP} interactions serve as the native entangling gates~\cite{1,kane}, our results clarify how structured environmental correlations govern the transition between classical and quantum memory regimes, thereby revealing their effect on decoherence dynamics and correlated errors \cite{NM,NM2}.}

\begin{figure}[t]
\centering
\includegraphics[width=\linewidth]{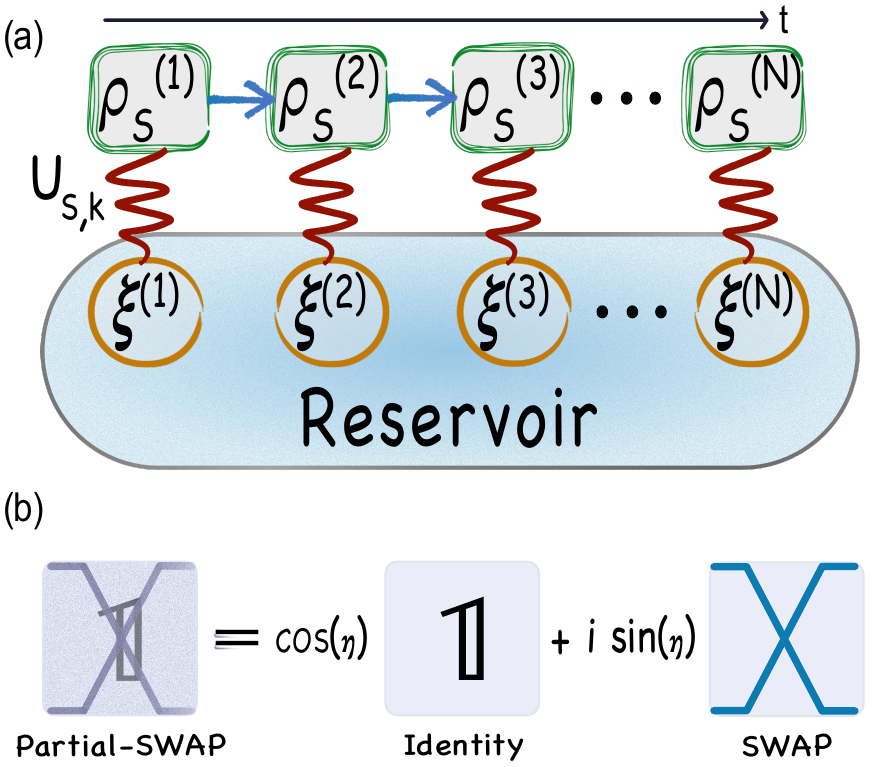}
\caption{{The Markovian dynamics setup.} Schematic of the quantum homogenizer, where a system qubit $S$ in state $\rho_{S}$ (green square) interacts sequentially ($k=1,...,N$) with the identically prepared $R$ qubits $\xi^{(k)}$ (orange circle). The partial-\texttt{SWAP} operator $\mathcal{U}_{{S},{k}}$, as in Eq. (\ref{eq:pswap}), mediates the unitary two-body operations. Similar to a noisy quantum channel \cite{alexander}, it is essentially a convex combination of the identity and \texttt{SWAP} gates.} 
\label{fig:figure1}
\end{figure}

\textit{Quantum homogenizer---}The homogenizer is a quantum machine composed of a set of $N$ qubits in state $\xi^{\otimes(k)}\in\mathcal{H}_{R}$ for $k \in {\{1,...,N \}}$, i.e., a reservoir, denoted as ${R}$. Where at the $k$th timestep, the system interacts with the $R$ state $\xi^{(k)} = \xi$. Generally, a system qubit $S$, initialized in some arbitrary state $\rho_{S}\in\mathcal{H}_{S}$, interacts with the $R$ qubits (one at a time) via the unitary partial-\texttt{SWAP}
\begin{equation}
\label{eq:pswap}
\mathcal{U}_{{S},{k}} = \text{cos} \left(\eta \right) \mathbb{1}_{{S}{k}} + i \,\text{sin} \left(\eta \right)\mathbb{S}_{{S}{k}} ,
\end{equation}
where $\mathbb{S}_{{S}{k}}$ is the \texttt{SWAP} operator between $S$ and the $k$th qubit in $R$ (i.e., the $k$th timesteps), and $\mathbb{1}_{{S}{k}}$ is the identity; with $\mathbb{S}\ket{\phi}\ket{\psi} = \ket{\psi}\ket{\phi}, \forall \ket{\phi},\ket{\psi}$, and $\mathbb{1}:\ket{\psi}\rightarrow\ket{\psi}$. Here, $\eta_k = \eta\in\left[0,\pi/2\right]$ denotes the coupling strength in the $k$th step, and is drawn from a Gaussian distribution with mean value $\eta$: the closer it is to $\pi/2$, the higher the probability is that the channel $\mathcal{U}_{{S},{k}}$ will act as the \texttt{SWAP} operation, Fig. \ref{fig:figure1}
Crucially, this time variance was recognized as an important quantum resource which endows the system with nonlinear, yet unitary, evolution, see, e.g., \cite{nonlinear,nonlinear2}. After $S$ interacts sequentially with all qubits in $R$, its state follows the recursive expression
\begin{equation}
\label{eq:densityoperator}
\rho^{(N)}_{{S}} = \text{tr}_{{R}}\left[\prod_{k=1}^{N} \mathcal{U}_{k}\left(\rho^{(0)}_{{S}}\otimes\xi^{\otimes (N)}\right) 
\right], 
\end{equation}
where $\mathcal{U}_{S, k}$ is given by Eq. (\ref{eq:pswap}) and $\prod$ is the discrete-time composition of the family of CPTP quantum maps $\{\mathcal{U}_{S, k}\}$. Applying Eq. (\ref{eq:pswap}) $N$ times and tracing over $R$ homogenizes any input state to $\xi$ in the asymptotic large-$N$ limit $\mathcal{U}_{{S},{k}}: \rho_{S}^{\left(N\right)}\rightarrow\xi$. The error of this transformation, i.e., the effectiveness of the machine, can be geometrically quantified as \cite{contract1}
\begin{equation}
\label{eq:distance2}
\|\mathcal{D}^{\left(N\right)}\|_{2} \equiv \| \xi^{\left(N\right)} - \rho_{{S}}^{\left(N\right)} \|_{2} , 
\end{equation}
where $\mathcal{D}$ is some distance measure, e.g., the $L_{2}$-norm \cite{contract2}. In the limit sense, the states converge perfectly
$ 
\lim_{N\rightarrow\infty}\|\mathcal{D}^{\left(N\right)}\|_{2}\rightarrow 0
$, which holds regardless of the initial states of $\xi$ and $\rho_{S}$ \cite{ziman,alexander}, and was even generalized in \cite{qudit} for the case of $d$-dimensional qudits (see also \cite{vlatko2, zz} for a recent numerical treatment of the role of coherence between the qubits in $R$). Notably, the convergence was proven in Ref. \cite{vlatko} as a necessary condition for the machine to be a \textit{constructor}.

\textit{Non-Markovian extensions---}We now consider two physical schemes both with the same system-reservoir, but different ancilla-ancilla interactions, enabling us to study the non-Markovian behavior. While deriving a general theoretical framework for the onset of quantum memory is a daunting task, here we show that both the structure of the $R$'s initial state and the nature of its interactions can decisively influence the memory type. By systematically varying these elements, we uncover surprising regimes, where quantum memory is either suppressed or activated.

In the first scheme, without loss of generality, let the first $R$ qubit $\xi^{(1)}_{C}$ be a control qubit, while $\xi^{(2)}$ and $\xi^{(3)}$ are two target qubits. After interacting with $S$ once through the partial-\texttt{SWAP} (\ref{eq:pswap}), the now entangled state reads
\begin{equation}
\label{eq:control}
\ket{\Psi}_{C,S}=\text{cos} \left(\eta\right)\ket{0}_{C}\ket{0}_{S} + \text{sin} \left(\eta\right)\ket{1}_{C}\ket{1}_{S} , 
\end{equation}
where $\ket{0}_{C}$ denotes the control qubit received no information from $\rho_{S}$ (i.e., $\mathcal{U}_{{S},{k}}$ acted as $\mathbb{1}_{{S}{k}}$), while $\ket{1}_{C}$ suggests that \textit{some} information was encoded. Then, depending on $\xi^{(1)}_{C}$, either the \texttt{SWAP} gate will be applied to $\xi^{(2)}$ and $\xi^{(3)}$, or they will be left unchanged. Formally, this describes the controlled-\texttt{SWAP} (Fredkin gate)~\cite{Note1} operation
\begin{equation}
\label{eq:cswap}
{\mathcal{C}_{k+1,k}} = \left(\ket{0}_{C}\bra{0}\otimes\mathbb{1}_{k+1, k} + \ket{1}_{C}\bra{1}\otimes\mathbb{S}_{k+1,k}\right) , 
\end{equation}
which imposes non-Markovianity on $R$, \textit{conditioned} on the partial-\texttt{SWAP} in Eq.~\eqref{eq:pswap} acting non-trivially. This way, prior to interacting with $S$ itself, the state of the $k$th qubit is dictated by the history of the partial-\texttt{SWAP}, and in particular, by the value of $\eta$. Hence, the strength of $\eta$ controls \textit{dynamically} the memory (i.e., the degree of non-Markovianity \cite{degree}) of $R$.

\revise{
In the second scheme, $S$ interacts with $\xi^{(1)}$ first and then the conditional ancilla-ancilla interaction (now between $\xi^{(1)}$ and $\xi^{(2)}$) takes place, via the CPTP map
\begin{equation}
\label{eq:ancillamap}
\mathcal{A}[\bullet] = \sqrt{p} [\bullet] + \sqrt{(1-p)} \mathbb{S}_{k+1,k} .
\end{equation}
Similarly, prior to the $k$th qubit interacting with $S$ itself, it will already be in a perturbed state. After the interaction between $\xi^{(1)}$ and $\xi^{(2)}$, the updated states are
\begin{equation}
\xi^{(i)}
= \cos^2\!\eta\,\xi^{(i)}
  + \sin^2\!\eta\,\xi^{(j)}
  + i\cos\eta\sin\eta\,[\xi^{(j)},\xi^{(i)}],
\end{equation}
where $(i,j)\in\{(1,2),(2,1)\}$ and $[\cdot,\cdot]$ is the commutator.}

Now we combine the system-ancilla (Eq. (\ref{eq:pswap})) and the ancilla-ancilla interactions (Eqs. (\ref{eq:cswap}, \ref{eq:ancillamap}), respectively) for the two schemes. Thus, we define the respective composite maps $\Phi:=\mathcal{C}\circ\mathcal{U}$ and $\Lambda:=\mathcal{A}\circ\mathcal{U}$. For the first scheme, at every timestep $t=k\tau$ (of fixed duration $\tau$) two operations are executed. For example, at $k=1$ we first apply the partial-\texttt{SWAP} between $S$ and $\xi^{(1)}$, followed by the controlled-\texttt{SWAP} between the next two qubits $\xi^{(2)}$ and $\xi^{(3)}$, with $\xi^{(1)}$ (as in Eq. (\ref{eq:control})) now serving as a control qubit. At $k=2$, $S$ interacts with $\xi^{(3)}$ and so on. Likewise for the second scheme: while at $k=1$ the first operation is identical, the second operation, given by Eq. (\ref{eq:ancillamap}), is applied between $\xi^{(1)}$ and $\xi^{(2)}$. The protocol continues by mere iteration, where at $k=2$, $S$ interacts with $\xi^{(2)}$. To formally construct $\Phi$ and $\Lambda$, let $\varphi_{S+R}^{(0)} := \rho_{S}^{(0)}\otimes\xi^{\otimes(N)}$ denote the initial state of the whole system.\\

\begin{figure}[t]
    \centering
    \includegraphics[width=\linewidth]{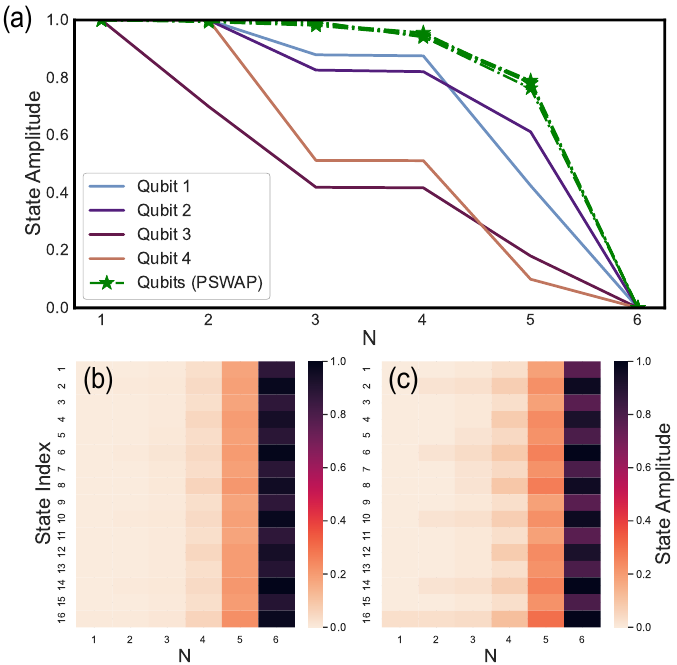}
    \caption{
        {Numerical simulation of the composite dynamics.}
        (a) Depiction of the non-Markovian evolution induced by the action of Eq.~(\ref{eq:composite}) on a bath of four qubits initialized in a product state. 
        The coupling strength $\eta \in [0, \pi/2]$ is drawn every $k$ from a Gaussian distribution and $N$ is the number of interactions. For performance comparison, we also show (in green) the same evolution but mediated by the traditional homogenizer instead. 
        (b)(c) Corresponding heatmap representations of the quantum state evolution via Eq.~(\ref{eq:pswap}) and Eq.~(\ref{eq:composite}), respectively. 
        Notably, we observe identical behavior as reported in Ref.~\cite{vlatko2,cswap2}: while the difference in convergence accuracy (i.e., effectiveness) between partial-\texttt{SWAP} (Markovian) and non-Markovian machines vanishes with increasing bath size, the trajectories (transient dynamics) of the evolving states in both scenarios differ significantly.}
    \label{fig:figure2}
\end{figure}

\textbf{Definition \hypertarget{definition}{1}.} \textit{For the composite maps $\Phi:=\mathcal{C}\circ\mathcal{U}$ and $\Lambda:=\mathcal{A}\circ\mathcal{U}$, given by Eqs. (\ref{eq:pswap}), (\ref{eq:cswap}), and (\ref{eq:ancillamap}),} $\Lambda^{(k)}, \Phi^{(k)} : \varphi_{S+R}^{(0)} \rightarrow \varphi_{S+R}^{(k)}$
\begin{equation}
\label{eq:composite}
\begin{aligned}
\Phi^{(1)}\!\left[\varphi_{S+R}^{(1)}\right] 
   &= \left(\mathcal{C}_{3,2} \circ \mathcal{U}_{S,1}\right)\!
      \left[\varphi_{S+R}^{(0)}\right] , \\[6pt]
\Lambda^{(1)}\!\left[\varphi_{S+R}^{(1)}\right] 
   &= \left(\mathcal{A}_{2,1} \circ \mathcal{U}_{S,1}\right)\!
      \left[\varphi_{S+R}^{(0)}\right],
\end{aligned}
\end{equation}
at $k=1$, and can be trivially extended in the large-$N$ limit. Ultimately, this characterizes both the efficiency and the trajectories of the state evolution of the machine, see Fig.~\ref{fig:figure2}. To quantify the degree of non-Markovianity, we use the distance measure (\ref{eq:distance2}) (a general version of the one proposed in \cite{degree}), which we herein define simply as \textit{loss of distinguishability} between the initial quantum states $\varphi_{S+R}^{(0)}$ under the action of the composite dynamics (\ref{eq:composite}) in the limit sense, see Fig.~\ref{fig:figure2}a. The question of how the coupling strength affects the states' convergence rate was examined numerically in \cite{alexander2} \cite{Note2}.

\begin{figure}[!t]
    \centering
    \includegraphics[width=\linewidth]{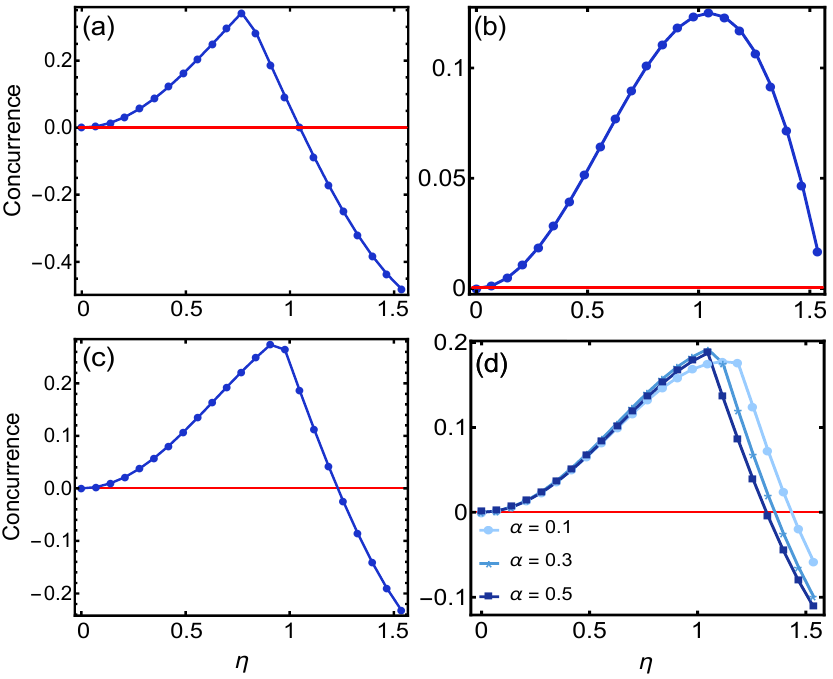}
    \caption{The entanglement differences $C^{\#}(\tilde{L}^{(k)}) - C(\tilde{M}^{(k)})$ and $C^{\#}(\tilde{L}^{(k)}) - C(\tilde{K}^{(k)})$ as a function of the coupling strength $\eta$ for $R$ initialized in (a) the Bell state, (b) the GHZ state $\ket{111}+\ket{000}$, (c) \textit{asymmetric} GHZ state $\ket{000}+\ket{101}$, and (d) perturbed GHZ state of the form $\sqrt{\alpha} \ket{100} + \sqrt{1-\alpha} $\textsc{GHZ}, with perturbation strength $\alpha=\{0.1, 0.3, 0.5\}$. In all cases, $\varphi_{S+R}^{(0)}$ evolves for $k=2$ steps under Eq.~(\ref{eq:composite}) for the two schemes, respectively. Numerically, for Bell, asymmetric, and perturbed GHZ state initializations (with $\alpha=0.5$), the concurrence witness Eq.~(\ref{eq:ineq}) becomes negative above a threshold of $\eta$.  In those regimes, quantum memory is needed. {The lower bound is $\eta = \pi/3$ saturated in the setting of (a).} For GHZ-type $R$, the concurrence remains positive and thus only classical memory suffices. However, for any non-zero value of the non-trivial perturbation $\alpha$, Eq.~(\ref{eq:ineq}) is satisfied, where higher values of $\alpha$ evoke quantum memory earlier.}
    \label{fig:figure3}
\end{figure}

\textit{Memory criteria---}We now analyze the type of memory necessary to realize Eq.~(\ref{eq:composite}) by using the criterion from \cite{memory} which, unlike existing methods (e.g., \cite{dof1, dof2}), is agnostic to the environmental degrees of freedom and only considers the local dynamics as a witness. The criterion makes use of the so-called \textit{entanglement of assistance} (proposed in \cite{assistance})
\begin{equation}
\label{eq:entofassist}
E^{\#}\left[\varphi_{S+R}\right] := \max_{\{{p}_{i},\ket{\psi_{i}}\}}\sum_{i}p_{i}E\left[\ket{\psi_{i}} \right] , 
\end{equation}
where $\varphi_{S+R}$ is a joint quantum state, $E\left[\bullet \right]$ is the 
concurrence (i.e., entanglement monotone \cite{concurrence}), and the entanglement is maximized over all pure state decompositions of $\varphi_{S+R}$. \revise{ Specifically, it is based on an inequality between the entanglement of assistance $C^{\#}$ and the entanglement of formation $C$ of the Choi states (denoted by a tilde) of the respective maps $C^{\#}(\tilde{\mathcal{U}}^{(k)})$, $C(\tilde{\Phi}^{(k)})$, and $C(\tilde{\Lambda}^{(k)})$. Note the Choi state \cite{choi} of a map is $\kappa[\bullet]=(\bullet \otimes\mathbb{1})\ket{\phi^{+}}\bra{\phi^{+}}$, where $\ket{\phi^{+}}=1/\sqrt{d}\sum_{i=0}^{d-1}\ket{i_{S}}\ket{i_{R}}$, $d$ is the system's dimension, and $\ket{i_{S,R}}$ forms an orthonormal basis. The unitaries (\ref{eq:composite}), however, are global (i.e., $\mathcal{U}^{(k)}, \Phi^{(k)}, \Lambda^{(k)}: \mathcal{H}_{S}\otimes\mathcal{H}_{R}\rightarrow \mathcal{H}_{S}\otimes\mathcal{H}_{R}$), while the Choi state is defined on $\mathcal{H}_{S}$ \cite{memory}. Therefore, we now define the corresponding \textit{local} CPTP map  $O^{(k)} $ on $S$ alone $O^{(k)} \in \{ L^{(k)},  M^{(k)}, K^{(k)} \}$, respectively, each acting as
\begin{equation}
O^{(k)}[\rho_S]
   = \text{tr}_R \big[\mathcal{X}^{(k)}(\rho_S\otimes\rho_R)\mathcal{X}^{\dagger(k)}\big],
\end{equation}
where $\mathcal{X}^{(k)}$ denotes the corresponding \textit{global} operation. The Choi states $\tilde{L}^{(k)}$, $\tilde{M}^{(k)}$, and $\tilde{K}^{(k)}$, characterizing the local dynamics, are then calculated; where given 
\begin{equation}
\label{eq:ineq}
C^{\#}(\tilde{L}^{(k)}) < C(\tilde{M}^{(k)}), \quad C^{\#}(\tilde{L}^{(k)}) < C(\tilde{K}^{(k)}),
\end{equation}
for the two schemes, the non-Markovian dynamics in Eq.~(\ref{eq:composite}) requires \textit{quantum} memory by virtue of Theorem 1 from \cite{memory}.} The easiest way to capture those effects is thereby to examine the evolution of $\rho_{S}$ across two consecutive timesteps under the composite dynamics of Eq.~(\ref{eq:composite}), and compare the corresponding monotones, Fig. \ref{fig:figure3}. That is, for the first scheme at $k=1$, the reduced dynamics is given by
$\rho_{S}^{(1)}  = \text{tr}_{{1}}\left[\mathcal{C}_{{3},{2}} \circ \mathcal{U}_{S,{1}} \left[\rho_{S}^{(0)} \right]\right]$, where the trace is taken over the first ancilla. Whereas at $k=2$, the evolution trivially reads
$\rho_{S}^{(2)} = \mathcal{U}_{S,3} \left[ \rho_{S}^{(1)} \right]$. Note that applying the second operation $\mathcal{C}_{{5},{4}}$ within $R$ is no longer needed as the evolution of $\rho_{S}$ is dictated entirely by the first map $\mathcal{U}$, Eq.~(\ref{eq:pswap}). Similarly for the second scheme, at $k=1$ we have $\rho_{S}^{(1)}  = \text{tr}_{{1}}\left[\mathcal{A}_{{2},{1}} \circ \mathcal{U}_{S,{1}} \left[\rho_{S}^{(0)} \right]\right]$, whereas $\rho_{S}^{(2)} = \mathcal{U}_{S,2} \left[ \rho_{S}^{(1)} \right]$ at $k=2$.

The controlled-\texttt{SWAP} plays a key role here: it simulates a \textit{conditional} memory mechanism, where it routes information from ancilla $k$ to $k+2$ (or keeps it where it is), so that when $S$ next collides with ancilla $k+2$, it interacts with a state that depends on its own past. This results in a bonafide information backflow without direct intermediate interaction with $S$. Alternatively, one can also initialize $R$ in a Bell state without controlled-\texttt{SWAP}.

\revise{\textit{Emergence of quantum memory---}}We aim to identify how different physically relevant conditions---such as $R$’s initial state and its entanglement structure---dictate the nature of the emerging memory from the reduced dynamics of $S$. To explore this, we systematically test broad classes of initial $R$ states, and compute $C^{\#}(\tilde{L}^{(k)}) - C(\tilde{M}^{(k)})$ and $C^{\#}(\tilde{L}^{(k)}) - C(\tilde{K}^{(k)})$, see Fig. \ref{fig:figure3}. Our results show that quantum memory can arise under a wide range of practical initializations. Importantly, we explore the optimal initialization of $R$. Below, we analyze the mechanisms behind that (see \cite{SM} for details).

Firstly, we initialize $R$ in the Bell state (Fig. \ref{fig:figure3}a) and the GHZ state (Fig. \ref{fig:figure3}b).
Although the GHZ state is genuinely multipartite entangled, its reduced two- or three-qubit marginals are separable and contain only classical correlations \cite{sep,sep2}. As a result, when $S$ interacts locally with such $R$, its dynamics can be simulated without accessing the global entanglement structure. Therefore, despite $R$ being initially entangled, the non-Markovian dynamics (\ref{eq:composite}) can be realized using classical memory. Conversely, when $R$ is initialized in the Bell state, the reduced two-qubit marginals retain quantum entanglement. 
\revise{
\begin{theorem}[informal]
\setcounter{theorem}{1}
\renewcommand{\thetheorem}{2}
Suppose system qubit $S$ is in initial state of the form $\alpha\ket{0}_{S} + \beta\ket{1}_{S}$, the $R$ qubits $\xi^{(1)}$ and $\xi^{(2)}$ are in arbitrary joint state, and the system-reservoir interaction is via Eq. (\ref{eq:pswap}). Then, the leading-order rate at which entanglement is transferred grows linearly in $\eta$.
\end{theorem}
Physically, entanglement ``leaks'' through the marginals despite that $S$ interacts only with one ancilla at a time; suggesting that even local interactions between $S$ and ancilla reveal genuine quantum correlations that cannot be reproduced by classical memory. Consequently, the dynamics of $S$ in this case requires access to quantum memory to be faithfully realized.} Then, we consider other realistic scenarios, in which some noise-induced errors are present. First, if a local perturbation is applied to the GHZ state—corresponding to, e.g., a single-qubit $X$ error—then classical memory is sufficient (not shown as the behavior is identical to Fig. \ref{fig:figure3}b). Interestingly, this suggests that the emergence of quantum memory is robust against local errors.
However, in the presence of non-trivial perturbations like: (i) coherent error in $R$ that yields \textit{asymmetric entanglement structure} or (ii) error of the form $\sqrt{\alpha} \ket{100} + \sqrt{1-\alpha} $\textsc{GHZ}, where $\alpha$ is the perturbation strength, we observe that indeed quantum memory is needed, Figs. \ref{fig:figure3}c and \ref{fig:figure3}d, respectively. Same behavior is also observed for the second scheme~\cite{Note3}.
\revise{
\begin{theorem}[informal]
\setcounter{theorem}{1}
\renewcommand{\thetheorem}{2}
Let $R$ contain additional qubits with asymmetric multipartite correlations. Then, sequential interactions suppress the leading-order terms, and entanglement between $S$ and $R$ appears at cubic order.
\end{theorem}}
The reason is as follows: perturbing the GHZ state introduces local entanglement into its marginals, making quantum correlations in $R$ accessible to $S$. As a result, non-Markovianity and quantum memory emerge, even though the unperturbed GHZ state required only classical memory. This indicates that if $R$ is structured, then even small imperfections can activate memory effects, underscoring the need for QEC schemes that account for correlated, history-dependent noise. 

\vspace{3pt}
\noindent 
\revise{
\textbf{Corollary 1.}
\textit{
Theorem~\ref{thm:GHZ_concurrence} implies an upper bound on the concurrence generation in a sequential interaction process with an intermediate decoherence channel.}}

\vspace{3pt}
\revise{
Furthermore, we observe that the Bell state provides maximally entangled two-qubit marginals, making quantum correlations in $R$ immediately accessible to $S$. 
\begin{proposition}
\setcounter{proposition}{1}
\renewcommand{\proposition}{2}
The two-qubit Bell marginal maximizes the off-diagonal coherence driving the first-order entanglement transfer from Theorem \ref{thm:thm1}.
\end{proposition}}

\revise{
This shows that  Bell marginal saturates the optimal entanglement transfer between $S$ and $R$, which is also observed numerically. In this case, weak system-reservoir coupling leads to dynamics that requires quantum memory, setting an activation lower bound for the onset of quantum memory, tied to the locality of the correlations. Equivalently, the above analytical result reveals the critical point at which a minimum interaction strength is required for quantum memory to emerge, regardless of how $S$ and $R$ are initialized. As numerically found in Fig.~\ref{fig:figure3}c, there exists a critical point $\eta = \pi/3$ which gives the lower bound of the interaction strength. In contrast, perturbed GHZ states contain weaker, more delocalized correlations, requiring stronger coupling to activate the memory, see Fig. \ref{fig:figure3}d.} 

\textit{Discussions---}We extended the standard quantum homogenizer to the non-Markovian regime by incorporating ancilla-ancilla interactions. While non-Markovian evolution naturally captures memory effects arising from system-environment correlations, a key challenge concerns identifying the nature of this memory, whether it is classical or genuinely quantum, and determining how it depends on the structure and correlations within $R$. Establishing this distinction is essential for understanding when non-Markovian dynamics can be simulated by classical resources and when inherently quantum memory is required. We addressed this question by using a recent local dynamics-based criterion as witness to quantum memory, and the concurrence as the entanglement measure. We identified physical conditions under which the non-Markovian dynamics requires genuine quantum memory, depending on the initialization and entanglement structure of $R$ and the propagation of non-classical correlations. That is, \revise{above a threshold, $R$ with \textit{asymmetric} entanglement (common in realistic noise models) or with certain perturbations require quantum memory, Figs. \ref{fig:figure3}c,d.} \revise{We support these observations with analytical results, which show the entanglement transfer for a broad class of $R$, and further establish the optimal initialization. Notably, this approach enables us to identify the activation lower bound for the onset of quantum memory which is based on the maximum entanglement transfer saturated by the Bell state. This is numerically investigated in Fig. \ref{fig:figure3}a}. We numerically demonstrated the state convergence properties (i.e., the effectiveness) of this machine (see Fig. \ref{fig:figure2}) and compared it against the traditional Markovian homogenizer. Our simulation corroborated earlier results \cite{vlatko2, cswap2} that analytically predicted the trajectories of the evolving states. 

Our results suggest that even local interactions between $S$ and $R$ reveal genuine quantum correlations that cannot be reproduced by classical memory. Specifically, we observe that for many practical models, even globally entangled states may not necessitate quantum memory, while certain coherent superpositions with separable marginals do. This indicates the emergence of memory is governed not just by the entanglement structure of $R$, but also by the propagation of non-classical correlations through it---a subtle, context-dependent property that invites further investigation. We proved that small imperfections in structured environments can activate memory effects even earlier, underscoring the need for QEC schemes that account for correlated, history-dependent noise. Our findings not only reveal when and how quantum memory becomes necessary to describe non-Markovian evolution, but also suggest a route towards  characterizing  and engineering memory in physical systems. The ability to tune memory activation via initial entanglement structure or reservoir interactions may be useful for designing noise models that better reflect real hardware behavior, or even for constructing memory-aware QEC schemes that exploit environmental coherence.

\textit{Acknowledgments---}V.V. thanks the Oxford Martin School, the John Templeton Foundation, the Gordon and Betty Moore Foundation, and the EPSRC (UK). J.S. would like to thank support from the Innovate UK (Project No.~10075020) and support through Schmidt Sciences, LLC.

\clearpage
\newpage

\widetext
\section*{Supplemental Materials}
\renewcommand{\addcontentsline}[3]{\oldaddcontentsline{#1}{#2}{#3}}

\tableofcontents

\setcounter{theorem}{0}
\label{app:appA}

\section{Analytical insights into the onset of quantum memory}

\subsection{Optimal reservoir initialization and saturation by the Bell marginals }

We show that the two-qubit marginal of a pair in $R$ initialized in the Bell state maximizes the leading-order entanglement (concurrence) generated between $S$ and the remote ancilla $\xi^{(2)}$. Without loss of generality, let the system qubit $S$ be initialized as $\ket{0}_S$, while the first two $R$ qubits $\xi^{(1)}$ and $\xi^{(2)}$ be in an arbitrary joint state $\rho_{1,2}$. The system-ancilla unitary is given by Eq. (\ref{eq:pswap}), where $\eta$ is assumed to be small. After $S$ interacts with $\xi^{(1)}$, the global state is

\begin{equation}
\rho_{g} 
= 
\mathcal{U}_{S,1}
\underbrace{\big(\ket{0}\bra{0}_S\otimes\rho_{1,2}\big)}_{:= \rho'}
\mathcal{U}_{S,1}^\dagger. 
\label{eq:global}
\end{equation}
Here, we study the perturbative scaling of the concurrence of the reduced state $\rho_{S,2}=\text{tr}_{1} \rho_{g}$.

\begin{theorem}[Linear transfer of entanglement between the reservoir and the system]
\setcounter{theorem}{0}
\renewcommand{\thetheorem}{1} 
\label{thm:thm1}
For the unitary partial-\texttt{SWAP} operator (\ref{eq:pswap}) between $S$ and $\xi^{(1)}$, to first order and for sufficiently small $\eta$, the concurrence of the reduced density matrix $\rho_{S,2}$ grows linearly in $|\kappa|$

\begin{equation}
E[\rho_{S,2}] = 2\eta|\kappa| + \mathcal{O}(\eta^2),
\label{eq:leading}
\end{equation}
where

\begin{equation}
\kappa:=\langle 01|\rho_{1,2}|10\rangle
\label{eq:kappa}
\end{equation}
is the off-diagonal coherence in the computational basis of $\xi^{(1)}\xi^{(2)}$ and encodes the amplitude exchange between $S$ and the ancilla degrees of freedom. 
\end{theorem}

\begin{proof}
The unitary operator (\ref{eq:pswap}) is expanded as a Taylor polynomial \cite{poly}

\begin{equation}
\mathcal{U}_{S,1} = \mathbb{1}_{S1} + i \eta \, \mathbb{S}_{S1} - \frac{\eta^2}{2} \mathbb{1}_{S1} + \mathcal{O}(\eta^3),
\end{equation}
where $\mathbb{S}$ is the standard \texttt{SWAP} between $S$ and $\xi^{(1)}$. This global operator acts as $\mathcal{U}_{S,1} \otimes \mathbb{1}_{2}$ on the full $SR$ space, and the evolved (\ref{eq:global}) is

\begin{equation}
\rho_{g} = \cos^{2}\eta\rho' - i \cos\eta\sin\eta[\mathbb{S}_{S1},\rho'] + \sin^{2}\eta\mathbb{S}_{S1}\rho'\mathbb{S}_{S1}.
\label{eq:rhoglobal}
\end{equation}
This gives the first-order term on expansion \cite{corr}

\begin{equation}
\rho_{g} = \rho' -i\eta[\mathbb{S}_{S1},\rho'] + \mathcal{O}(\eta^{2}),
\end{equation}
while to second order it reads

\begin{equation}
\rho_{g} = \rho' - i\eta [\mathbb{S}_{S1}, \rho'] + \eta^2 (\mathbb{S}_{S1} \rho' \mathbb{S}_{S1} - \rho') + \mathcal{O}(\eta^3),
\end{equation}
where the term $\eta^2 (\mathbb{S}_{S1} \rho' \mathbb{S}_{S1} - \rho')$ accounts for the state exchange that populates $R$'s subspace at $\mathcal{O}(\eta^{2})$. Note that at zeroth order, the reduced state $\rho_{S,2} = \text{tr}_{1} \rho_{g}$ reads $\rho_{S,2}^{0} = \ket{0}\bra{0}_S \otimes \rho_2$, where $\rho_2 = \text{tr}_1 \rho_{1,2}$, which has no $S$-$\xi^{(2)}$ coherence.
From Eq. (\ref{eq:rhoglobal}), tracing out $\xi^{(1)}$ and evaluating the $S$-$\xi^{(2)}$ off-diagonal element gives

\begin{equation}
\begin{aligned}
z &= \langle 0_S 1_2 | \rho_{S,2} | 1_S 0_2 \rangle \\[4pt]
& = -i \cos\eta \sin\eta \, \kappa + \mathcal{O}(\eta^{2}),
\label{eq:offdiagonal}
\end{aligned}
\end{equation}
which arises from the anti-commutator-like behavior of the \texttt{SWAP} operator, and yields $z = -i\eta \, \kappa + \mathcal{O}(\eta^{2})$, with $\kappa$ defined as Eq. (\ref{eq:kappa}). Here, the contribution of the first-order term $-i\eta \, \text{tr}_1 [\mathbb{S}_{S1}, \rho']$ is

\begin{equation}
z^{1} = -i\eta \langle 0_S 1_2 | \text{tr}_1 (\mathbb{S}_{S1} \rho' - \rho' \mathbb{S}_{S1}) | 1_S 0_2 \rangle,
\end{equation}
where $\text{tr}_1 \rho' \mathbb{S}_{S1} = \ket{0}\bra{0}_S \otimes \text{tr}_1 (\rho_{1,2} \mathbb{S}_{S1})$ since $\rho' = \ket{0}\bra{0}_S \otimes \rho_{1,2}$, and the leading term comes from $\text{tr}_1 \mathbb{S}_{S1} \rho'$. Therefore, since the diagonal entries change only at $\mathcal{O}(\eta^{2})$, $\rho_{S,2}$ is an X-state of the form

\begin{equation}
\rho_{S,2} = \begin{pmatrix}
a & 0 & 0 & 0 \\
0 & c & z & 0 \\
0 & z^* & e & 0 \\
0 & 0 & 0 & f
\end{pmatrix},
\end{equation}
with $a = p_0 - \mathcal{O}(\eta^2)$, $c = p_1 - \mathcal{O}(\eta^2)$, $e,f = \mathcal{O}(\eta^2)$, and $z = -i\eta \, \kappa + \mathcal{O}(\eta^2)$ per Eq. (\ref{eq:offdiagonal}), where $p_0 = \langle 0|\rho_2|0 \rangle$ and $p_1 = \langle 1|\rho_2|1 \rangle$ denote the populations. The concurrence \cite{concc} of $\rho_{S,2}$ is

\begin{equation}
E[\rho_{S,2}] = 2 \max\left(0, |z| - \sqrt{a f}, 0 - \sqrt{c e}\right) + \mathcal{O}(\eta^2),
\end{equation}
where the second term is negative for small $\eta$, and \text{max} selects the dominant off-diagonal term. Thus, $E[\rho_{S,2}] = 2 |z| + \mathcal{O}(\eta^2) = 2 \eta |\kappa| + \mathcal{O}(\eta^2)$; cf. Eq. (\ref{eq:leading}).
\end{proof}

We now turn to the optimality in the off-diagonal coherence between the $R$ qubits.
\setcounter{proposition}{0}

\begin{proposition}[Optimal reservoir initialization]
\label{thm:thm2}
For any two-qubit density matrix $\rho_{1,2}$, the upper bound of the off-diagonal matrix element

\begin{equation}
|\kappa| = |\langle 01|\rho_{1,2}|10\rangle| \le \frac{1}{2}
\label{eq:bound}
\end{equation}
is saturated if and only if $\rho_{1,2}$ is a projector of the form $c_{1}\ket{01}+c_{2}\ket{10}$ with equal populations $|c_{1}|^{2}=|c_{2}|^{2}=1/2$.
\end{proposition}
Consequently, the leading-order concurrence (\ref{eq:leading}) is maximized by the two-qubit Bell marginal. 

\begin{proof}
The positivity \cite{pos} of $\rho_{1,2}$ implies the $2\times2$ submatrix on $\{|01\rangle, |10\rangle\}$ 

\begin{equation}
M =
\begin{pmatrix}
p_{01} & \kappa^{*} \\
\kappa & p_{10}
\end{pmatrix}
\label{eq:submatrix}
\end{equation}
is also positive semidefinite, $p_{01} = \bra{01}\rho_{1,2}\ket{01}$, $p_{10} = \bra{10}\rho_{1,2}\ket{10}$ are diagonal populations, and $\det(M)\geq 0$. Hence

\begin{equation}
p_{01} p_{10} - |\kappa|^2 \ge 0 ,
\end{equation}
yielding $|\kappa|^2 \le p_{01} p_{10}$ and implying $M$ is rank-one \cite{pos2}. Then it is easy to see that

\begin{equation}
p_{01} p_{10} \le \left( \frac{p_{01} + p_{10}}{2} \right)^2 \le \frac{1}{4} ,
\end{equation}
where $p_{01} + p_{10} \le \text{tr} \rho_{1,2} = 1$. Here, $\kappa = p_{01} = p_{10} = 1/2$ and $p_{00} = p_{11} = 0$ is achieved by the Bell state. 
\end{proof}
Let's now look at multipartite state initialization.
\setcounter{theorem}{1}
\begin{theorem}[Multipartite entanglement transfer]
\label{thm:GHZ_concurrence}
\setcounter{theorem}{2}
\renewcommand{\thetheorem}{2} 
For system qubit \(S\) initialized in \(|0\rangle_S\) and a three-qubit reservoir \(R = \{\xi^{(1)},\xi^{(2)},\xi^{(3)}\}\), 
prepared in the asymmetric GHZ state

\begin{equation}
\label{eq:productstate}
\rho_{1,2,3} = \frac{1}{\sqrt{2}}\big(|000\rangle + |101\rangle\big),
\end{equation}
let $S$ interact sequentially with \(\xi^{(k)}\) (for $k=1,2,3$) via the partial-\texttt{SWAP} operator (\ref{eq:pswap}). Then, the concurrence between \(S\) and \(\xi^{(3)}\)

\begin{equation}
E[\rho_{S,3}] = \sin^3\!\eta\,\cos\!\eta\,\big(1-\sin^2\!\eta\big)
\label{eq:C_GHZ_exact}
\end{equation}
scales at cubic order in $\eta$.
\end{theorem}

\begin{proof}
The initial global $SR$ state is now \cite{multipart}

\begin{equation}
\rho_{g}
= \ket{0}\bra{0}_S \otimes \rho_{1,2,3}
= \frac{1}{\sqrt2}\big(|0000\rangle + |0101\rangle\big),
\end{equation}
where after applying Eq. (\ref{eq:pswap}) successively, the final state becomes

\begin{equation}
\label{eq:finalstate}
\begin{aligned}
\ket{\Psi_g} = \frac{1}{\sqrt2}\Big[
& e^{-3 i\eta}\ket{0000}
+ e^{-i\eta}\cos^2\eta\ket{0101} - i e^{-i\eta}\sin\eta\cos\eta \big(\ket{1100}+\ket{1001}\big) \\
& - \sin^2\eta\cos\eta\ket{0011}
+ i\sin^3\eta\ket{1010}
\Big].
\end{aligned}
\end{equation}
Projecting Eq. (\ref{eq:finalstate}) onto the $S\xi^{(3)}$ subspace, and tracing out \(\xi^{(1)}\) and \(\xi^{(2)}\), the reduced two-qubit state $\rho_{S,3}=\text{tr}_{1,2}\rho_{g}$ is an X-state \cite{rev} that reads

\begin{equation}
\rho_{S,3} =
\begin{pmatrix}
a & 0 & 0 & b\\
0 & c' & z & 0\\
0 & z^* & e & 0\\
b^* & 0 & 0 & f
\end{pmatrix},
\end{equation}
where $a = 1/2$ and

\begin{equation}
\label{eq:elements}
\begin{aligned}
c' &= \frac{\cos^2 \eta}{2}\big(\cos^2 \eta + \sin^4 \eta\big), \\[4pt]
e &= \frac{\sin^2 \eta}{2}\big(\cos^2 \eta + \sin^4 \eta\big), \\[4pt]
f &= \frac{\sin^2 \eta \cos^2 \eta}{2}, \\[4pt]
b &= \frac{i}{2}\sin\eta \cos\eta \, e^{-2i\eta}, \\[4pt]
z &= \frac{i}{2}\sin\eta \cos\eta \big(\cos^2 \eta + \sin^4 \eta\big).
\end{aligned}
\end{equation}
Here, the concurrence is

\begin{equation}
\label{eq:ghzconcurrence}
E[\rho_{S,3}] = 2\max\{0,\,|b|-\sqrt{c' e},\,|z|-\sqrt{a f}\},
\end{equation}
where substituting (\ref{eq:elements}) gives

\begin{align}
|b| - \sqrt{c'e} &= \frac{1}{2} \sin^3 \eta \cos \eta (1 - \sin^2 \eta) > 0, \nonumber\\
|z| - \sqrt{a f} &= -\frac{1}{2} \sin^3 \eta \cos \eta (1 - \sin^2 \eta) < 0.
\end{align}
Evidently, the two channels $|b|$ and $|z|$ are competing, where the coherence between $\ket{01}$ and $\ket{10}$ does not contribute as it equals $\sqrt{c'e}$. Eq. (\ref{eq:ghzconcurrence}) then becomes

\begin{equation}
E[\rho_{S,3}] 
= 2(|b|-\sqrt{c'e})
= \sin^3\!\eta\,\cos\!\eta\,(1-\sin^2\!\eta),
\end{equation}
where expanding for finite \(\eta\) gives \(E[\rho_{S,3}] = \eta^3 + \mathcal{O}(\eta^5)\).
\end{proof}

\setcounter{corollary}{0}
\begin{corollary}
As a consequence of Theorem~\ref{thm:GHZ_concurrence}, the growth of the concurrence $E[\rho_{S,3}]$ is upper bounded by the maximum coherence $\rho_{1,3}^{\mathrm{max}}$ for any sequential process with an intermediate CPTP map $\mathcal{E}_{2}$. 
\end{corollary} 
We can think of the process described in Theorem~\ref{thm:GHZ_concurrence} as the action of a composite quantum channel on the initial $SR$ state. Explicitly

\begin{equation}
\rho_{S,3}
= 
\text{tr}_{1,2}\!\left[
\left(\mathcal{U}_{S,3}\circ\mathcal{E}_{2}\circ\mathcal{U}_{S,1}\right)
\big(\rho_{S}\otimes\rho_{R}\big)
\right],
\end{equation}
where $\rho_{R}=\rho_{1,3}\otimes\rho_{2}$, and $\mathcal{E}_{2}$ is a CPTP map describing the intermediate collision with $\xi^{(2)}$. Formally

\begin{equation}
\mathcal{E}_{2}(\rho)
= 
\sum_{j}
\big(P_{j}\otimes\mathbb{1}_{1,3}\big)
\rho
\big(P_{j}^{\dagger}\otimes\mathbb{1}_{1,3}\big),
\end{equation}
where $\sum_{j}P_{j}^{\dagger}P_{j}=\mathbb{1}$ denotes the Kraus operators representing local decoherence (including amplitude damping, phase damping, or thermal reset, see \cite{NM2, rev2}) induced by the uncorrelated qubit $\xi^{(2)}$. Since the resulting concurrence is a non-linear function of $\rho_{R}$, we can write

\begin{equation}
E[\rho_{S,3}] = E[\mathcal{F}(\rho_{R})],
\end{equation}
where $\mathcal{F}$ is a linear superoperator representing the channel composition and acting on the initial $R$ state.  
For initialization as in Eq. (\ref{eq:productstate}), the intermediate map $\mathcal{E}_{2}$ effectively acts as a decoherence channel that suppresses the leading-order (linear and quadratic) coherence terms, resulting in the cubic scaling $E[\rho_{S,3}] \sim \mathcal{O}(\eta^3)$. From the monotonicity of entanglement \cite{monot} it follows that for any $\rho_{1,3}$ within this product structure

\begin{equation} 
E[\mathcal{F}(\rho_{1,3} \otimes \rho_2)] \leq E[\mathcal{F}(\rho_{1,3}^{\mathrm{max}} \otimes \rho_2)], 
\end{equation}
where $\rho_{1,3}^{\mathrm{max}}=|\Phi^{+}\rangle\langle\Phi^{+}|$ maximizes the coherence bound from Proposition~\ref{thm:thm2}, and the concurrence is non-increasing under $\mathcal{E}_{2}$. Hence, Theorem~\ref{thm:GHZ_concurrence} provides an upper bound of concurrence generation for sequential processes with an intermediate decoherence channel. In a way, the result represents the \textit{surviving} quantum memory. This higher-order, non-linear term corresponds to a ``scrambled'' resource that persists after the leading-order correlations are erased by $\mathcal{E}_2$. The cubic onset can thereby be used as a diagnostic and distinct signature of a non-Markovian quantum memory suppressed by structural decoherence.

\subsection{Comments on the generalization to arbitrary system states}

\subsubsection{State-dependent oscillations}

Our results extend beyond the initialization considered above. The linear-order entanglement growth in Theorem \ref{thm:thm1} is a feature of Eq. (\ref{eq:pswap}) which is a coherent entanglement swapping operation, driven by $R$'s coherence $\kappa$. Due to the linearity of the unitary evolution, the final reduced state $\rho_{S,k}$ contains direct and interference terms

\begin{equation}
\rho_{S,k}
= |\alpha|^{2}\rho^{(0)}
+ |\beta|^{2}\rho^{(1)}
+ \left( \alpha\beta^{*}\rho^{(\mathrm{int})} + \text{h.c.}\right),
\end{equation}
where $\rho^{(0,1)}$ are the density matrices resulting from the $\ket{0}$ and $\ket{1}$ evolution pathways, respectively, and $\rho^{(\text{int})}$ denotes the cross-terms from the quantum interference. Since the matrix elements of $\mathcal{U}$ depend trigonometrically on $\eta$ \cite{alexander} (via $\cos\eta$ and $\sin\eta$), $\rho^{(\text{int})}$ introduces a state-dependent oscillatory modulation to the concurrence. However, this does not change the scaling, as the leading-order contribution remains linear, given by the $R$ marginal $\rho_{1,2}$
\begin{equation}
E[\rho_{S,2}] = 2\eta \left| \mathcal{F}(\alpha, \beta, \rho_{1,2}) \right| + \mathcal{O}(\eta^2),
\end{equation}
where $\mathcal{F} := \sum_{ij} c_{ij}(\alpha,\beta) \langle i|\rho_{1,2}|j\rangle$. Here, the indices $i,j$ run over the computational basis of the two-qubit $R$ marginal $\rho_{1,2}$, and the coefficients $c_{ij}(\alpha,\beta)$ arise from the perturbative expansion of the conjugate interaction channel. Thus, $\mathcal{F}$ characterises the effective entanglement-swapping strength: it is a linear functional of the reservoir coherences that determines the prefactor of the leading-order $\mathcal{O}(\eta)$ entanglement generation.

\subsubsection{Reduction by local basis rotation}

The previous subsection explains the arbitrary system initialization based on physical argument of the concurrence, in which different states bring some oscillatory behavior. Here, we give another explanation based on local basis transformations. Any arbitrary pure $S$ initialization of the form $\rho_S=\alpha\ket{0}_S+\beta\ket{1}_S$
can be generated from the state $\ket{0}_S$ by a local unitary $V_S$, such that $\rho_S=V_S\ket{0}_S$.
Where for any number of sequential interactions via Eq. (\ref{eq:pswap}), we can define the total evolution $\mathcal{U}_{\text{tot}}$ on $S \otimes R$. The reduced system state is then given by
\begin{align}
\rho_{S,k}
&= \text{tr}_R\Big[\mathcal{U}_{\text{tot}}\big(V_S\ket{0}\bra{0}_S V_S^\dagger\otimes\rho_R\big)\mathcal{U}_{\text{tot}}^\dagger\Big]\notag\\
&= V_S \left( \text{tr}_R\Big[\widetilde{\mathcal{U}}_{\text{tot}}\big(\ket{0}\bra{0}_S\otimes\rho_R\big)\widetilde{\mathcal{U}}_{\text{tot}}^\dagger \Big] \right)V_S^\dagger,
\end{align}
where $\widetilde{\mathcal{U}}_{\text{tot}} =(V_S^\dagger\otimes\mathbb{1}_R)\mathcal{U}_{\text{tot}}(V_S\otimes\mathbb{1}_R)$ is the conjugate evolution operator. Since the concurrence is invariant under local unitary transformations on $S$, the problem is then equivalent to calculating the concurrence generated by $\ket{0}_S$ under $\widetilde{\mathcal{U}}_{\text{tot}}$. We can expand $\widetilde{\mathcal{U}}_{\text{tot}}$ perturbatively in $\eta$, where the conjugate \texttt{SWAP} is
\begin{equation}
\widetilde{\mathbb{S}}_{Sj}=(V_S^\dagger\otimes\mathbb{1}_j)\mathbb{S}_{Sj}(V_S\otimes\mathbb{1}_j).
\end{equation}
Crucially, $\widetilde{\mathbb{S}}_{Sj}$ acts on the $R$ index $j$ exactly as the original $\mathbb{S}_{Sj}$, but couples them to rotated bases on $S$. Consequently, the first-order contribution to $\rho_{S,k}$ is proportional to the commutator with $\widetilde{\mathbb{S}}_{Sj}$. Because the $R$ part of the operator is unchanged, the trace samples the same $R$ matrix elements as in the $\ket{0}_{S}$ case. The $S$ rotation $V_S$ merely modifies the numerical coefficients of these terms. Therefore, the leading-order scaling established for $\ket{0}_S$ is preserved.

\subsection{The emergence of quantum memory across different time blocks}

In the $N$-step interaction limit of the quantum homogenizer, the reduced state of the system qubit $S$ converges asymptotically to the fixed point $\xi$, but the underlying dynamics remains non-Markovian at every finite step\footnote{To be more precise, each step represents a unit block of the dynamics, where we apply a sequence of system-reservoir coupling operations (rather than just one or two operators) as dictated by the two schemes in the main text.}. This is because the local map at step $k{+}1$ depends on coherent information injected into the adjacent $R$ cells at step $k$. By Theorem \ref{thm:thm1}, the entanglement that $S$ transfers to the next neighboring ancilla scales as $E \approx 2\eta|\kappa|$, where $\kappa=\langle 01|\rho_{j,j+1}|10\rangle$ is the off-diagonal coherence of the two-qubit marginal. Thus, whenever the adjacent marginal exhibits $|\kappa|>0$ (as in Bell-like or dimerised reservoirs), each collision regenerates $\mathcal{O}(\eta)$ coherent backflow, showing that quantum memory and non-Markovianity persist for all finite $N$. 


When considering variable block sizes and non-adjacent step, the key point is that the emergence of quantum memory at the $i^{th}$ block guarantees the emergence of quantum memory at the $j^{th}$ block ($j > i$), provided that all the blocks are identical. {Note that: (i) this is the case for the quantum homogenizer model\footnote{It is often the case that the system-reservoir coupling repeats periodically in time.} and (ii) the emergence of quantum memory is independent of the initialization of $S$ (as shown above, different $S$ initialization\footnote{The reservoir is initialized identically in different blocks and the variation between blocks can be regarded as arising solely from the system’s initial state.} only affects the magnitude but the sign is unchanged and thus still allows us to identify the onset of quantum memory).} Considering {block-dependent initialization of $R$,} different block sizes, or non-subsequent timesteps dilute the generality of our result as in this model-dependent scenarios, the criteria we employ should be adjusted for the situation of interest.

For more general reservoir states entangled across $k >2$ qubits, the behavior is still dictated by the two-qubit marginals when employing the witness criteria from \cite{memory} over two consecutive timesteps. If the $k$-partite state has separable marginals (e.g., GHZ-type), the leading-order $\mathcal{O}(\eta)$ contribution is suppressed and memory emerges only at higher order, as in Theorem \ref{thm:GHZ_concurrence} where the scaling becomes $\mathcal{O}(\eta^{3})$. This pushes the onset of quantum memory to larger $\eta$. Conversely, if the $k$-partite $R$ retains short-range entanglement—such as in valence-bond or dimerised structures—then $|\kappa|$ remains large on adjacent pairs, and the onset of quantum memory occurs at the same low values of $\eta$ as in the Bell-marginal case.

\section{Applications to solid-state quantum computing}

In solid-state devices, the Heisenberg exchange interaction, equivalent to the partial-\texttt{SWAP} (\ref{eq:pswap}), is the native entangling gate \cite{kane}. Recall that when $R$ is initially in Bell or perturbed GHZ state, quantum memory is needed even though the underlying interactions are strictly local, Fig.~\ref{fig:figure3}. Physically, this means that implementing even two-local exchange-based gates or composite operations like the controlled-\texttt{SWAP} (which naturally arises in multi-qubit exchange scenarios) can require quantum memory, especially in the presence of hidden correlations in the environment \cite{hidden}. This finding has direct relevance for exchange-based models: although the Heisenberg exchange is optimal in the gate count \cite{solidstate}, it does not inherently suppress correlated errors. Our results show that even weak, structured correlations that are naturally present in solid-state environments---such as those induced by spin-orbit coupling or charge noise in quantum dots---can evoke quantum memory. In these cases, even high-fidelity local operations may be subject to non-Markovian errors \cite{hidden}. This is because non‑Markovian noise often contains a coherent part that evolves on comparable or longer timescales than the gate sequence \cite{coh}. As the environment retains memory, the induced error phases accumulate coherently across successive gates rather than averaging out, thereby producing history-dependent errors, Theorem \ref{thm:GHZ_concurrence}. Here, $\eta$ defines the effective interaction strength per gate. We show that when $\eta \geq \pi/3$, the $R$’s correlations become quantum-coherent, and classical noise models break down. Identifying this threshold can provide a quantitative design rule for distinguishing between classical and quantum memory regimes in real devices. This moreover provides a classification of quantum channels implementability via repeated interactions. While memoryless CPTP maps (e.g., depolarizing and amplitude damping) can be simulated with classical memory, structured environments with asymmetric or coherent entanglement require quantum memory.

Meanwhile, for surface codes in exchange-based systems, it is known that temporally correlated errors on syndrome qubits \cite{err1} can lead to suboptimal decoding \cite{codes,codes2}. Our results imply that these correlations may, in fact, originate not only classically but also from quantum correlations in the environment, which cannot be eliminated trivially \cite{err2}. Recognizing this can inform both hardware-level mitigation strategies (e.g., dynamical decoupling) and adaptations like correlation-aware decoders, ultimately improving fault tolerance in solid-state computing. Thus, our results provide strong evidence that these features play a decisive role in shaping memory behavior.

\section{Operating regimes}We show the model belongs to a class of collision models~\cite{cmclass} that are governed by CPTP evolution and can easily interpolate between Markovian and non-Markovian regimes. Clearly, the composite structure of Eq. (\ref{eq:composite}) indicates departure from the Markovian semigroup \cite{5} which is more restricted as it is characterized by a family of time-homogeneous one-parameter dynamical maps that are typically used in simpler (memoryless) collision models. Interestingly, given Definition \hyperlink{definition}{1}, and from Refs. \cite{semigroup, kernel}, $\{\Phi^{(k)}\}$ forms a family of contractive time-\textit{inhomogeneous} dynamical maps (i.e., a series of jump operations that satisfy memory kernel functions \cite{semigroup}). We can, therefore, write the evolution of $\varphi_{S+R}^{(0)}$ (Eq. (\ref{eq:composite})) as a sum of operations

\begin{equation}
\label{eq:sum}
\begin{aligned}
\varphi_{S+R}^{(N)} = &\left(1-p\right)\sum_{k \in \textrm{odd}}^{N-1} p^{k-1} \, \mathcal{U}_{S,k} \left[ \varphi_{S+R}^{(N-k)} \right] + p^{N-1} \, \prod_{k \in \textrm{even}}^{N-1} \mathcal{C}_{k+2,k+1} \left[ \varphi_{S+R}^{(0)} \right],
\end{aligned}
\end{equation}
where $p^{k}$ (dictated by $\eta$ at every timestep $k$) denotes the probability of non-trivial operation between $S$ and $R$ ancilla. Since the application of the controlled-\texttt{SWAP} is \textit{conditioned} on the partial-\texttt{SWAP} operator (which only acts on one $R$ qubit at a time), we can use it as a witness of whether the second operation (given by Eq. (\ref{eq:cswap})) will induce non-Markovianity. In a way, the state $\rho_{S}$ is not only necessary but \textit{sufficient} in order to know how $S$ has evolved after $N$ timesteps. That is, given $\rho_{S}$ is only affected by the partial-\texttt{SWAP} (cf. Eq. (\ref{eq:composite})), we can thereby have an identical expression to Eq. (\ref{eq:sum}) but just in terms of the evolution of $\rho_{S}$ (see, e.g., \cite{semigroup})

\begin{equation}
\label{eq:rhoS}
\begin{aligned}
\rho_{S}^{(N)} &= (1-p) \sum_{k=1}^{N-1} p^{k-1} 
    \mathcal{M}^{(k)}\!\left[\rho_{S}^{(N-k)}\right] + p^{N-1} \mathcal{M}^{(N)}\left[\rho_{S}^{(0)}\right] ,
\end{aligned}
\end{equation}
where $\mathcal{M}$ is a CPTP map, corresponding to Eq. (\ref{eq:pswap}); see \cite{alexander2,alexander} for details. Generally speaking, this evolution describes the alternative situation of weak repetitive interactions between $S$ and a single qubit, which is known to be strongly non-Markovian according to the established criteria \cite{criteria1, criteria2}, and lead to similar state convergence as in Fig.~\ref{fig:figure2}a. From Eq. (\ref{eq:rhoS}), it is now clear that we can interpolate between the Markovian and non-Markovian regimes by simply controlling the coupling strength $\eta$ (which corresponds to the probability $p$, as discussed earlier; cf. Eq. (\ref{eq:sum})). Note that analogous non-Markovian quantum chain structure was further discussed in \cite{qchain} and experimentally tested using linear optics in \cite{optics}.


\end{document}